\documentclass[12pt]{article} 

\usepackage{epsfig} 
\usepackage{amsfonts}
\usepackage{amssymb}
\usepackage{amsmath}  
\usepackage{amsbsy}  
\usepackage{wasysym}
\usepackage{color}

\catcode`\@=11 
\@addtoreset{equation}{section} 
\def\theequation{\thesection.\arabic{equation}} 
 
\@twosidetrue 

\catcode`\@=11  
\def\section{\@startsection{section}{1}{\z@}{3.5ex plus 1ex minus 
.2ex}{2.3ex plus .2ex}{\large\bf}}

\def\thesection{\arabic{section}} 
\def\thesubsection{\arabic{section}.\arabic{subsection}} 
\def\thesubsubsection{\arabic{section}.\arabic{subsection}.\arabic{subsubsection}} 
 
\def\appendix{\setcounter{section}{0} 
 \def\thesection{\Alph{section}} 
 \def\theequation{\Alph{section}.\arabic{equation}} 
\def\thesubsection{\Alph{section}.\arabic{subsection}} 
\def\thesubsubsection{\Alph{section}.\arabic{subsection}.\arabic{subsubsection}} 
 
\def\section{\@startsection{section}{1}{\z@}{3.5ex plus 1ex minus 
   .2ex}{2.3ex plus .2ex}{\large\bf}} } 
 
\newcount\minutes 
\newcount\scratch 
 
\def\timestamp{%
\scratch=\time 
\divide\scratch by 60 
\edef\hours{\the\scratch} 
\multiply\scratch by 60 
\minutes=\time 
\advance\minutes by -\scratch 
---$\,$\hours:\null 
\ifnum\minutes< 10 0\fi 
\the\minutes}

\begin{document} 
\begin{titlepage} 
\nopagebreak 
{\flushright{ 
        \begin{minipage}{5cm}
         KA--TP--09--2006  \\            
        {\tt hep-ph/0609075}\hfill \\ 
        \end{minipage}        } 
 
} 
\vfill 
\begin{center} 
{\LARGE \bf 
 \baselineskip 0.5cm 
Anomalous Higgs boson couplings in vector boson fusion at the CERN LHC
} 
\vskip 0.5cm  
{\large   
V. Hankele, G. Kl\"amke, D. Zeppenfeld 
}   
\vskip .2cm
{\it Institut f\"ur Theoretische Physik, 
        Universit\"at Karlsruhe, P.O.Box 6980, 76128 Karlsruhe, Germany}
\vskip .2cm
{and}
\vskip .2cm
{\large T. Figy}
\vskip .2cm
{\it Institute of Particle Physics Phenomenology, University of Durham, Durham, DH1 3LE, UK}  
 
 \vskip 
1.3cm     
\end{center} 
 
\nopagebreak 
\begin{abstract}
Deviations from SM expectations in the Higgs sector can be
parameterized by an effective Lagrangian. The corresponding anomalous
couplings have been implemented in a Monte Carlo program for Higgs
production in vector boson fusion, at NLO QCD accuracy. It allows to
study anomalous coupling effects for production and decay of the Higgs
boson. We analyze deviations allowed by LEP data and study a 
new azimuthal angle variable which directly measures the interference
between CP-even, CP-odd and SM couplings.

\end{abstract} 
\vfill 
\hfill 
\vfill 

\end{titlepage} 
\newpage               
%
%
\section{Introduction}
\label{sec:intro}

At the LHC, the second most copious source for a standard model (SM)  
type Higgs boson is expected to be the vector boson fusion (VBF)
channel, i.e. electroweak processes of the type 
$qq\to qqH$~\cite{Spira:1997dg,Djouadi:2005gi}.  While the 
production cross section from gluon fusion is larger, VBF has the 
advantage of a richer kinematic structure with two forward tagging
jets which result from the scattered quarks. The characteristic
distributions of these forward and backward tagging jets allow for
significant reduction of backgrounds which should result in fairly
clean samples of signal events. These samples can then be used to
measure properties of the Higgs boson, in particular its couplings to
gauge bosons and fermions. This includes the magnitude of the 
couplings~\cite{Zeppenfeld:2000td} but also the tensor structure of the
$HVV$ vertex ($V=W,Z$)~\cite{Plehn:2001nj}.

\begin{figure}[hbt] 
\centerline{ 
\epsfig{figure=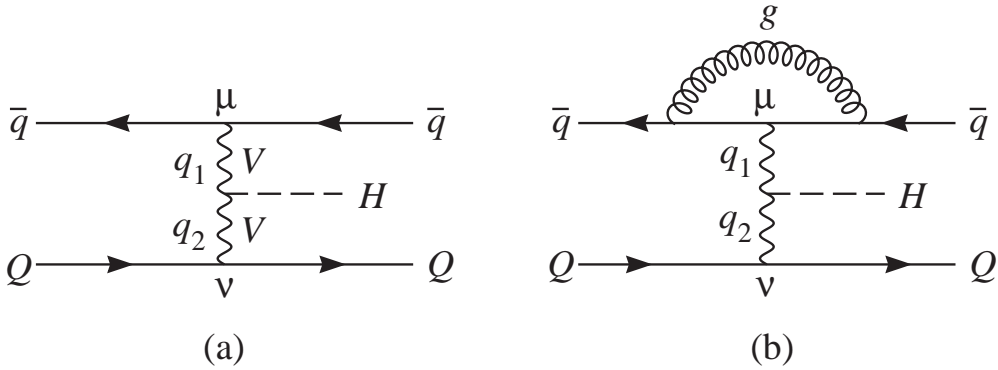,width=0.8\textwidth,clip=}
} 
\caption{ 
\label{fig:feyn2} 
Feynman graphs contributing to $\bar qQ\to \bar qQH$ at
(a) tree level and (b) including virtual corrections to the upper quark line.
The momentum labels and Lorentz indices for the internal weak bosons 
correspond to the vertex function of Eq.~(\ref{eq:vertex}).
}
\end{figure} 

The total cross section for Higgs production via VBF has been known at
NLO QCD accuracy since the early nineties~\cite{Han:1992hr}. More
recently, the NLO corrections to distributions have also been
calculated, and are available in the form of NLO parton level Monte Carlo
programs for a SM Higgs boson~\cite{Figy:2003nv,Berger:2004pc} and also for
a general tensor structure of the $HVV$ coupling~\cite{Figy:2004pt}, as
depicted in Fig.~\ref{fig:feyn2}. The most general tensor
structure of the $HVV$ vertex which can contribute to VBF in the
massless quark limit can be written as 
\begin{equation}
T^{\mu \nu}(q_{1},q_{2}) = a_1(q_1,q_2)\; g^{\mu \nu} + 
a_2(q_1,q_2)\; [q_{1}\cdot q_{2} g^{\mu \nu} - q_{2}^{\mu}q_{1}^{\nu}] + 
a_3(q_1,q_2)\; \varepsilon^{\mu \nu \rho \sigma} q_{1 \rho} q_{2 \sigma} \;.
\label{eq:vertex}
\end{equation} 
A constant $a_1$ (with $a_2=0=a_3$) represents the SM case while
sizable form factors $a_2$ and/or $a_3$ would represent new physics,
induced, for example, by a heavy particle loop. Such new physics effects
would, 
of course, not only change Higgs production cross sections but also the 
decay rates and branching ratios for $H\to VV$ ($V=W,Z,\gamma$).
A convenient starting point for a consistent 
treatment of such correlated effects on production and decay 
is a description by an effective Lagrangian, which is constructed out of
EW gauge fields and the SM Higgs doublet field in a $SU(2)\times U(1)$  
gauge invariant way. The corresponding
operators~\cite{Buchmuller:1985jz,Hagiwara:1993ck}  have been considered
in the past for describing new physics contributions to Higgs
physics, see e.g.~\cite{Hagiwara:1993qt,Eboli:1999pt,Manohar:2006gz}. 
They are described in Section~\ref{sec:Leff}. We have now incorporated 
these effects on Higgs decays into our VBFNLO program, which allows
to simulate general VBF processes with NLO QCD
accuracy~\cite{Figy:2003nv,Oleari:2003tc}. One purpose of the present
paper is to make available this new simulation tool for Higgs signals at
the LHC. We give a brief description of our program in
Section~\ref{sec:tools}. 

The remainder of this paper analyzes signals of anomalous Higgs
couplings in VBF processes at the LHC.
Sizable effective $H\gamma\gamma$ and $HZ\gamma$ couplings of a sufficiently
light Higgs boson would have led to $e^+e^-\to H\gamma$ events at LEP
and are, hence, tightly constrained~\cite{Achard:2004kn}. These
constraints limit the maximal LHC signals. In Section~\ref{sec:res} we
analyze the implications for total Higgs production cross sections in
VBF, we consider the effect on Higgs branching ratios, and we use our
simulation tools to find distributions which are sensitive to small
interference effects between the three different tensor structures of
Eq.~(\ref{eq:vertex}). Of particular interest here are distributions
of the two (anti-)quark jets in VBF events. These forward and
backward tagging jets are characteristic features of the VBF process
and their distributions and correlations can be exploited to reveal
information on the tensor structure of the $HVV$ vertex, independent of
the Higgs decay mode. In 
Ref.~\cite{Plehn:2001nj} it was shown that the absolute value of
the azimuthal angle between the two tagging jets can distinguish between
the three different choices of tensor structures in Eq.~(\ref{eq:vertex}).
However, interference effects between the CP-even
coupling $a_2$ and the CP-odd coupling $a_3$ cancel in this
distribution, and results for $|a_2|=|a_3|$ are very close to SM
predictions. Here we show that the azimuthal angle can be defined in
such a way that also its sign can be determined at the LHC and that this
new angle exhibits the interference between $a_2$ and
$a_3$. Indeed, the ratio of the two form factors can be directly 
measured via the position of the minimum
of the azimuthal angle distribution. Such interference effects would
signal CP-violation in the Higgs sector. Conclusions are given in
Section~\ref{sec:concl}. 

\section{Effective Lagrangian and anomalous couplings}
\label{sec:Leff}

We are concerned with deviations in the $HV_1V_2$ couplings
($V_i=W,Z,\gamma$) from SM predictions, and, more generally, with new
physics effects in the bosonic sector of the SM. A
model-independent description of such effects is provided by an effective
Lagrangian approach, where, in order to preserve the successful SM predictions 
for $W$ and $Z$ interactions with fermions, the SM $SU(2)\times U(1)$
gauge symmetry is taken as exact, albeit spontaneously broken.
It is, hence, required for all higher
dimensional operators in the effective Lagrangian~\cite{Buchmuller:1985jz},
\begin{align} \label{eq:leff}
{\cal L}_{eff} = {\cal L}_{SM} + \sum_i \frac{f_i^{(6)}}{\Lambda^2}
{\cal O}_i^{(6)} + ... \;.
\end{align}
With a scalar doublet field giving rise to the Higgs boson, 
an even number of covariant derivatives and of Higgs doublet fields is
required which leaves operators of even dimensionality only.
The relevant operators for
our discussion are four CP-even and three CP-odd
operators of dimension 6, namely
\begin{align} \label{eq:Operators}
\begin{split}
&{\cal O}_{BB} = \Phi^+\  \hat{B}_{\mu\nu} \hat{B}^{\mu\nu}\  \Phi
\hspace{2.2cm} {\cal O}_{\tilde{B}B} = \Phi^+\  \hat{\tilde{B}}_{\mu\nu}
\hat{B}^{\mu\nu}\ \Phi\\
&{\cal O}_{WW} = \Phi^+ \ \hat{W}_{\mu\nu} \hat{W}^{\mu\nu}\  \Phi
\hspace{1.8cm} {\cal O}_{\tilde{W}W} = \Phi^+ \ \hat{\tilde{W}}_{\mu\nu}
\hat{W}^{\mu\nu}\ \Phi\\
&{\cal O}_{B} = (D_\mu \Phi)^+\ \hat{B}^{\mu\nu}\ (D_\nu \Phi)
\hspace{1.4cm} {\cal O}_{\tilde{B}} = (D_\mu \Phi)^+\
\hat{\tilde{B}}^{\mu\nu}\ (D_\nu \Phi)\\
&{\cal O}_{W} = (D_\mu \Phi)^+ \hat{W}^{\mu\nu}\ (D_\nu \Phi).
\end{split}
\end{align}
In this formula the covariant derivative, $D_\mu$, the field strength tensors,
$\hat{B}_{\mu\nu}$ and $\hat{W}_{\mu\nu}$, of the W and B gauge fields
and their dual ones are given by:
\begin{align}
\begin{split}
&D_\mu=\partial_\mu + \frac{i}{2} \ g'\ B_\mu + i \ g\
\frac{\sigma^a}{2}\ W_\mu^a,\\
&\hat{B}_{\mu\nu} + \hat{W}_{\mu\nu} = i\ \frac{g'}{2} \ B_{\mu\nu} + i
\ \frac{g}{2} \ \sigma^a \ W_{\mu\nu}^a = [D_\mu, D_\nu]\\
&\tilde{V}_{\mu\nu} = \frac{1}{2} \ \epsilon_{\mu\nu\rho\sigma} \
V^{\rho\sigma}, \qquad (V = B, \ W).
\end{split}
\end{align}
Two other operators of dimension 6, ${\cal O}_{\Phi,1} = (D_\mu \Phi)^+ \
 \Phi \Phi^+ \ (D^\mu \Phi)$ and ${\cal O}_{BW} = \Phi^+ \
\hat{B}_{\mu\nu} \ \hat{W}^{\mu\nu}\ \Phi$, contribute to anomalous $HVV$
couplings, but have already been constrained strongly by electroweak
high precision measurements~\cite{Hagiwara:1993ck,Erler:2004nh} and will
be neglected in the following.

The notation we refer to is the one used by the L3
collaboration~\cite{Achard:2004kn} with the effective Lagrangian
\begin{align} \label{eq:LHVV}
\begin{split}
{\cal L}_{eff}^{(6)}\ &=\ g_{H\gamma\gamma}\ H A_{\mu\nu} A^{\mu\nu}\
+ \ g_{HZ\gamma}^{(1)}\ A_{\mu\nu} Z^\mu \partial^\nu H + \
g_{HZ\gamma}^{(2)}\ H A_{\mu\nu} Z^{\mu\nu}\\
&+ \ g_{HZZ}^{(1)}\ Z_{\mu\nu} Z^\mu \partial^\nu H + \ g_{HZZ}^{(2)}\ H
Z_{\mu\nu} Z^{\mu\nu} + \ g_{HWW}^{(2)}\ H W_{\mu\nu}^+ W_-^{\mu\nu}\\
&+\ g_{HWW}^{(1)}\ (W_{\mu\nu}^+ W_-^\mu \partial^\nu H\ +\ W_{\mu\nu}^-
W_+^\mu \partial^\nu H \ ) + \ \tilde{g}_{H\gamma\gamma}\ H
\tilde{A}_{\mu\nu} A^{\mu\nu}\\
&+ \ \tilde{g}_{HZ\gamma} \ H \tilde{A}_{\mu\nu} Z^{\mu\nu} + \
\tilde{g}_{HZZ}\ H \tilde{Z}_{\mu\nu} Z^{\mu\nu} + \
\tilde{g}_{HWW}^{(2)}\ H \tilde{W}_{\mu\nu}^+ W_-^{\mu\nu}.
\end{split}
\end{align}

The $g_{HVV}$ are coefficients of the CP-even and the
$\tilde{g}_{HVV}$ the ones of the CP-odd
operators. $g_{HZ\gamma}^{(1)}$, $g_{HZZ}^{(1)}$ and $g_{HWW}^{(1)}$ can
be parameterized using the well known coefficients of the anomalous triple
gauge boson couplings, $\triangle \kappa_\gamma$ and $\triangle
g_1^Z$ ~\cite{Hagiwara:1986vm}. They are already highly restricted by a
combination of the measurements of the four LEP
collaborations~\cite{unknown:2004qh}.
The CP-odd coefficients $\tilde{g}_{HZ\gamma}$  and $\tilde{g}_{HZZ}$
depend on the parameter $\tilde{\kappa}_\gamma$, which has also been
constrained in the past by LEP data~\cite{Schael:2004tq}.
The remaining coefficients depend on two parameters,
\begin{align}
d = -\frac{m_W^2}{\Lambda^2}\ f_{WW},  \hspace{3cm} d_B = -
\frac{m_W^2}{\Lambda^2}\ \frac{\sin^2{\theta_w}}{\cos^2{\theta_w}}\ f_{BB}
\end{align}
for the CP-even and completely analogous on $\tilde{d}$ and
$\tilde{d}_B$ for the CP-odd couplings.
\begin{align} \label{eq:ghVV}
\begin{split}
&g_{H\gamma\gamma} = \frac{g}{2\  m_W}\ (d\ \sin^2{\theta_w} + d_B\
\cos^2{\theta_w}) \hspace{5cm} \\
&g_{HZ\gamma}^{(2)} = \frac{g}{2\ m_W}\ \sin{2 \theta_w} \ (d-d_B)\\
&g_{HZZ}^{(2)} = \frac{g}{2\ m_W}\ (d\ \cos^2{\theta_w} + d_B\
\sin^2{\theta_w})\\
&g_{HWW}^{(2)} = \frac{g}{m_W}\ d.
\end{split}
\end{align}
The best constraints for the parameters $d$ and $d_B$ come from the L3
collaboration. They are really bounds on $g_{H\gamma\gamma}$ and
$g_{HZ\gamma}^{(2)}$, which have been derived using the L3 bounds 
on the Higgs partial widths $\Gamma(H\to\gamma\gamma)$ and 
$\Gamma(H\to Z\gamma)$ of Fig. 7(a) in Ref.~\cite{Achard:2004kn}. 
For these couplings and also for $d$ and $d_B$, the strongest
constraints are from the unsuccessful search for the 
process $e^+ e^- \to H\gamma \to \gamma\gamma\gamma$ via photon
or $Z$-boson exchange in Higgs production.
Direct bounds on
$g_{HZZ}^{(2)}$ can only be derived from the process $e^+ e^- \to H Z$
 and, thus, for Higgs masses below $\approx 114$~GeV.

\begin{table}[bt]
  \caption{Approximate direct 95 \% CL bounds for
    $g_{H\gamma\gamma}$ and $g^{(2)}_{HZ\gamma}$ in $[\mathrm{TeV}^{-1}]$.
    They are derived from the partial decay width bounds of 
    Ref.~\cite{Achard:2004kn}. }
\begin{center}
\begin{tabular}{c||c|c}
Higgs mass & $g_{H\gamma\gamma}$ & $g^{(2)}_{HZ\gamma}$\\
\hline
\hline
& &\\
120 GeV & $[ -0.17, 0.17 ]$ & $[-1.24, 1.24 ]$\\
& &\\
\hline
& &\\
140 GeV & $[ -0.24, 0.24 ]$ & $[ -1.35, 1.35 ]$\\
& &\\
\end{tabular}
\end{center}
\label{tab:directbounds}
\end{table}

Since the parameters $d$ and $d_B$ also appear in the other coefficients
of Eq.~(\ref{eq:ghVV}), the L3 constraints  can be used to estimate upper 
bounds for $g_{HZ\gamma}^{(2)}$, $g_{HZZ}^{(2)}$ and
$g_{HWW}^{(2)}$. The best indirect bounds are summarized in
Table~\ref{tab:ghVVd} and Table~\ref{tab:ghVVdb}.
The parameter $d_B$ is stronger constrained than
$d$ because in $g_{H\gamma\gamma}$ they appear in the combination $d_B
\cdot \cos^2{\theta_w}$ and $d \cdot \sin^2{\theta_w}$. 
Since $\cos^2{\theta_w} \approx 3 \cdot \sin^2{\theta_w}$ this means 
that the best bounds on $d_B$ are about
three times better than for $d$. 
Below we shall use the maximal values of the $g_{HVV}$ induced by 
a pure $d$ coupling, i.e. the ${\cal O}_{WW}$ operator, to estimate maximal
allowed deviations from the SM which might arise from anomalous $HVV$ 
couplings.

\begin{table}[bt]
  \caption{The best indirect 95 \% CL constraints of the L3
    collaboration~\cite{Achard:2004kn} for $g_{H\gamma\gamma}$,
    $g^{(2)}_{HZ\gamma}$, $g^{(2)}_{HZZ}$ and $g^{(2)}_{HWW}$ in
    $[\mathrm{TeV}^{-1}]$, assuming $SU(2) \times U(1)$ invariant
    dimension 6 effective couplings with $d_B = 0$.}
\begin{center}
\begin{tabular}{c||c|c|c|c}
Higgs mass & $g_{H\gamma\gamma}$ & $g^{(2)}_{HZ\gamma}$ &
$g^{(2)}_{HZZ}$ & $g^{(2)}_{HWW}$\\
\hline
\hline
& & & &\\
120 GeV & $[ -0.16, 0.16 ]$ & $[ -0.59, 0.59]$ & $[-0.54, 0.54 ]$ & $[
-1.41, 1.41]$\\
& & & &\\
\hline
& & & &\\
140 GeV & $[ -0.24, 0.22 ]$ & $[ -0.86, 0.79]$ & $[-0.78, 0.72 ]$ & $[
-2.04, 1.88]$\\
& & & &\\
\hline
& & & &\\
150 GeV & $[ -0.31, 0.29 ]$ & $[ -1.12, 1.01 ]$ & $[ -1.02, 0.96 ]$ & $[
-2.66, 2.51 ]$\\
& & & &\\
\hline
& & & &\\
160 GeV & $[ -0.47, 0.37 ]$ & $[ -1.72, 1.35 ]$ & $[ -1.57, 1.23 ]$ & $[
-4.07, 3.21 ]$\\
& & & &\\
\end{tabular}
\end{center}
\label{tab:ghVVd}
\end{table}

\begin{table}[bt]
  \caption{The best indirect 95 \% CL constraints of the L3
    collaboration~\cite{Achard:2004kn} for $g_{H\gamma\gamma}$,
    $g^{(2)}_{HZ\gamma}$, $g^{(2)}_{HZZ}$ and $g^{(2)}_{HWW}$ in
    $[\mathrm{TeV}^{-1}]$, assuming $SU(2) \times U(1)$ invariant
    dimension 6 effective couplings with $d = 0$.}
\begin{center}
\begin{tabular}{c||c|c|c}
Higgs mass & $g_{H\gamma\gamma}$ & $g^{(2)}_{HZ\gamma}$ &
$g^{(2)}_{HZZ}$\\
\hline
\hline
& & &\\
120 GeV & $[ -0.18, 0.18 ]$ & $[ -0.20, 0.20]$ & $[-0.05, 0.05 ]$\\
& & &\\
\hline
& & &\\
140 GeV & $[ -0.24, 0.24 ]$ & $[ -0.26, 0.26]$ & $[-0.07, 0.07 ]$\\
& & &\\
\hline
& & &\\
150 GeV & $[ -0.27, 0.27 ]$ & $[ -0.30, 0.30 ]$ & $[ -0.08, 0.08 ]$\\
& & &\\
\hline
& & &\\
160 GeV & $[ -0.39, 0.39 ]$ & $[ -0.43, 0.43 ]$ & $[ -0.12, 0.12 ]$\\
& & &\\
\end{tabular}
\end{center}
\label{tab:ghVVdb}
\end{table}

Although the L3 collaboration does not give the corresponding values for
the CP-odd couplings they are implicitly constrained by existing data,
because CP-even and CP-odd couplings give identical contributions to the
differential cross section for $e^+ e^- \to H \gamma$:

\begin{align}
\frac{d \sigma}{d \Omega} \ ( e^+ e^- \to H \gamma ) =
\frac{\alpha^2}{8}\ (1 - \frac{m_H^2}{s})^3\ (1 + \cos^2{\theta})\ (F(s)
+ \tilde{F}(s))
\end{align}
with
\begin{align*}
F(s) &= 4\ \frac{g^2_{H\gamma\gamma}}{e^2} + \frac{1}{e^2}\
(2 \ g_{HZ\gamma}^{(2)})^2\ \frac{1}{64\ \cos^2{\theta_w} \sin^2{\theta_w}}\\
&\cdot \big( (1 - 4 \sin^2{\theta_w})^2 + 1 \big)\
\big(\frac{s}{s-m_Z^2}\big)^2 + \frac{1}{e^2}\ g_{H\gamma\gamma}\
g_{HZ\gamma}^{(2)}\\
&\cdot \frac{1}{\cos{\theta_w} \sin{\theta_w}}\ (1 - 4
\sin^2{\theta_w})\ \big(\frac{s}{s-m_Z^2}\big)
\end{align*}
and similarly for $\tilde{F}(s)$, with 
$g_{H\gamma\gamma} \to \tilde{g}_{H\gamma\gamma},
\ g^{(2)}_{HZ\gamma} \to \tilde{g}_{HZ\gamma}$.

%
%
\section{Calculational tools}
\label{sec:tools}

In this analysis we use a fully flexible Monte-Carlo program to
determine effects of the anomalous couplings at the LHC that are
compatible with existing LEP data. The program, part of the VBFNLO
package~\cite{Figy:2003nv,Oleari:2003tc} is similar to the one used in
Ref.~\cite{Figy:2004pt} but also includes anomalous
couplings in the Higgs boson decay modes $H\to\gamma\gamma$, 
$H\to Z\gamma$, $H\to ZZ$, and $H\to WW$. The subsequent SM decays of $W$ and
$Z$ bosons to fermion-antifermion pairs are implemented including full
finite width effects, i.e. the weak bosons are allowed to be off-shell.
In the presence of anomalous couplings to both $Z$ and photon,
interference effects between $\gamma$ and $Z$ exchange graphs are
included for the decay. Anomalous couplings can be entered in
the format given by Ref.~\cite{Figy:2004pt}, which is similar to the one
of Eq.~(\ref{eq:vertex}), by using the parameters $d$ and $d_B$,  or by using 
the coefficients $f_i$ of the dimension 6 operators or their CP-odd
analogs.

For the Higgs production processes, anomalous $H\gamma\gamma$ and
$HZ\gamma$ couplings introduce new photon fusion contributions. Their
interference with the SM $ZZ$ fusion process is included in the code
and NLO QCD corrections to the resulting full anomalous matrix elements
are implemented along the lines described in
Refs.~\cite{Figy:2003nv,Figy:2004pt}. At present, the program provides 
full NLO cross sections at parton level, for arbitrary
distributions. However, $Hjj$ and $Hjjj$ cross sections at LO QCD can be
generated as unweighted events which are then interfaced with parton
shower programs by making use of the Les Houches standard
interface~\cite{Boos:2001cv}. 

Anomalous couplings in production and decay give rise to $1/\Lambda^4$
terms in the naive amplitude and hence one might worry whether
consistency requires the inclusion of dimension 8 operators. However,
the decay effects are automatically unitarized by including the
anomalous couplings also in the calculation of the total Higgs decay
width which enters the Higgs boson propagator. As long as anomalous
couplings do lead to a narrow Higgs width, the observable rate
factorizes into Higgs production cross section times decay branching
fraction, $\sigma\times B$, and each factor is properly described by the
inclusion of dimension 6 operators only, as long as one probes
momentum transfers well below the scale of new physics, $\Lambda$. 
For the light Higgs boson masses considered here 
(in the 100 to 200 GeV range) the assumption $m_H\ll \Lambda$ is well
motivated. For Higgs production, however, momentum transfers
$|q_i^2|\approx \Lambda^2$ might be reached in the available phase space.
As a consequence, form factor effects, as indicated in
Eq.~(\ref{eq:vertex}) should be investigated. The VBFNLO code supports
form factors of the form 
\begin{equation}
a_i(q_1,q_2) = a_i(0,0)\; \frac{M^2}{|q_1^2|+M^2}\; \frac{M^2}{|q_2^2|+M^2}
\end{equation}
and 
\begin{equation}
a_i(q_1,q_2) = a_i(0,0)\; 2M^2 C_0(q_1,q_2;M)
\end{equation}
for the coefficients of the tensors in Eq.~(\ref{eq:vertex}).
Here $M$ corresponds to the mass scale of new physics (e.g. the mass of
a heavy particle going around in a loop) and $C_0$ is the usual scalar
loop integral for triangle graphs.

%
%
\section{Predictions for the LHC}
\label{sec:res}

The anomalous $HV_1V_2$ couplings defined in Section~\ref{sec:Leff}  
will affect cross sections for Higgs boson production via VBF at the
LHC. For a detailed analysis one would have to differentiate between 
different Higgs decay channels. Here we aim at a general view
of possible changes to the observable VBF cross sections, within the 
typical cuts which will be applied to enhance the Higgs signal.
We require the presence of two hard tagging jets which are defined as 
the two highest $p_T$ jets of the event. These tagging jets are then 
required to be widely separated in rapidity, in opposite detector 
hemispheres, and they must have a large invariant mass. Following 
Ref.~\cite{Kauer:2000hi} we set these cuts at
\begin{align}
\begin{split}
p_{T_{Jet}} \geq 20 \ \mathrm{GeV}, \qquad |y_j| \leq 4.5, \qquad
y_{j_{1}} \cdot y_{j_{2}} < 0,\\
m_{jj} > 600 \ \mathrm{GeV}, \qquad \Delta y_{tags} = |y_{j_{1}}
-y_{j_{2}}| \geq 4.2, \\
y_{j,min} + 0.6 < \eta_{l_{1,2}} < y _{j,max} -0.6.
\end{split}
\label{eq:cuts}
\end{align}
The final cut restricts the pseudo-rapidity range of the Higgs decay 
products, here dubbed ``leptons'' and assumed to be light, to lie 
between the jet definition cones of the two tagging jets. It is intended
to roughly simulate the requirement that the Higgs decay products be 
central. All cross sections to be presented below have been generated 
at LO and with the cuts of Eq.~(\ref{eq:cuts}). The LO approximation 
is adequate here since the NLO corrections are small, typically below 5\%.

\begin{figure}[htb]
\centerline{
\includegraphics[height=7.2cm, width=8.1cm]{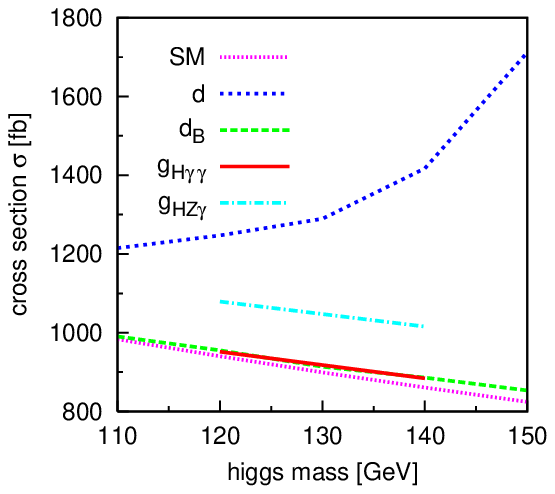} \hspace*{0.2cm}
\includegraphics[height=7.2cm, width=8.6cm]{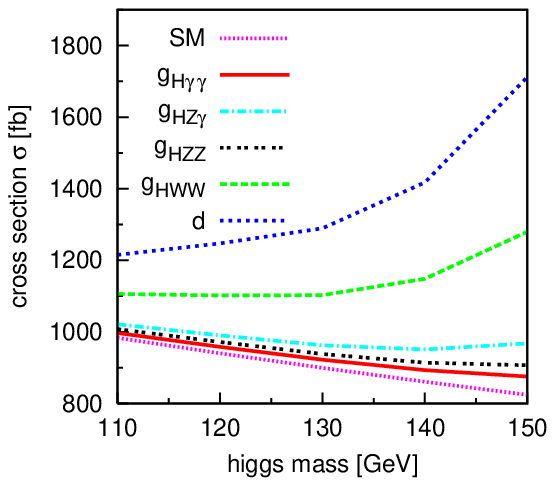}
}
\caption[]{\label{fig:WQ1}
 Production cross section of a Higgs Boson with anomalous
 couplings in addition to the SM $HVV$ interactions. 
{\it Left:} The upper bounds for $g_{H\gamma\gamma}$ and
 $g_{HZ\gamma}^{(2)}$ given in Table~\ref{tab:directbounds}  and the 
upper 95\% CL bounds for $d$ and $d_B$, given by the 
L3 collaboration, are used.
{\it Right:} The upper bound for $d$ is used for the different couplings.
}
\end{figure}

On the one hand Fig.~\ref{fig:WQ1} shows the LHC production cross
section for a Higgs boson with only a single additional anomalous
coupling of Eq.~(\ref{eq:LHVV}) and on the other hand production cross
sections for the parameters $d$ or $d_B$ are shown. On the left side, 
couplings 
$g_{H\gamma\gamma}$ and $g_{HZ\gamma}^{(2)}$ have been used which 
saturate the 95\% CL bounds of Table~\ref{tab:directbounds}. 
$g_{H\gamma\gamma}$ is already tightly constrained by the absence of an
$e^+e^-\to H\gamma,\;H\to \gamma\gamma$ signal at LEP. Only minor deviations
from the SM are allowed in the VBF Higgs production cross section due to this 
coupling alone. The direct limit on a pure $g_{HZ\gamma}^{(2)}$ coupling is 
weaker, since the more difficult $H\to Z\gamma$ partial width would have
been enhanced at LEP,  and this coupling still allows for significant
deviations from SM expectations in the Higgs production cross section at
the LHC. The parameter $d_B$ only leads to small deviations from the SM and can
be neglected, whereas a large enhancement due to the parameter $d$ is
still possible.

On the right hand side of Fig.~\ref{fig:WQ1} the indirectly determined
constraints on the anomalous couplings of Table~\ref{tab:ghVVd} and,
thus, for $d_B = 0$ have been used and cross sections for individual
anomalous couplings at the 95\% CL limit are shown. $g_{HZ\gamma}^{(2)}$ and
$g_{HZZ}^{(2)}$ already have strong bounds and do not lead to
significant deviations from the SM. However, $g_{HWW}^{(2)}$ as given in
Eq.~(\ref{eq:ghVV}) only depends on $d$ and is mainly responsible for the 
large enhancement of the cross section.

In Higgs decay, the effects of $g_{H\gamma\gamma}$ and
$g_{HZ\gamma}^{(2)}$ are the most important ones, because the $H\gamma\gamma$
and $HZ\gamma$ couplings in the SM only appear at one loop level and are
therefore very small. With the anomalous couplings the partial decay
widths of $H \to  \gamma \gamma, Z \gamma$ can be enhanced by several
orders of magnitude and even become the dominant decay channels. Since
in the SM the couplings of the Higgs boson to $W$ and $Z$ bosons appear
already at tree level, the effects of anomalies in the $HZZ$ and 
$HWW$ couplings are not so strong.

\begin{figure}[tb]
\centerline{
\includegraphics[height=7.2cm, width=7.5cm]{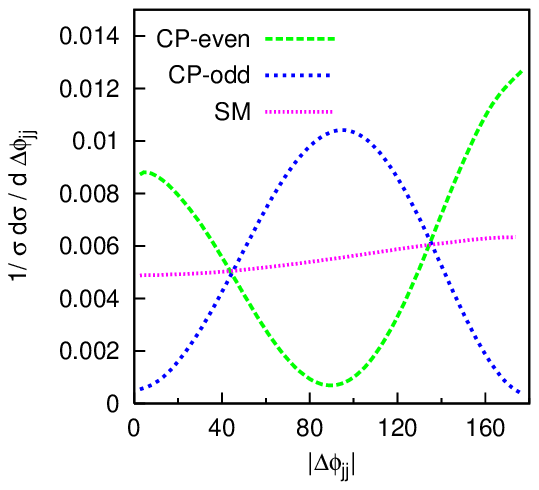} \hspace*{0.2cm}
\includegraphics[height=7.2cm, width=7.8cm]{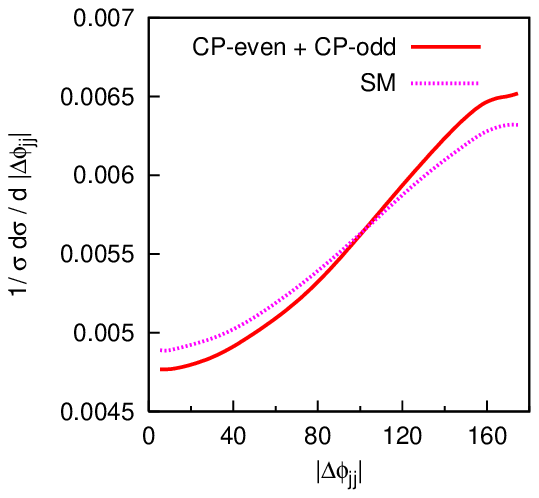}
}

\caption[]{\label{fig:Azimuthal1}
 Normalized azimuthal angle correlations of the two tagging
jets in VBF as in Ref.~\cite{Plehn:2001nj} for a Higgs mass of 120 GeV
and $d$ = 0.18 or, respectively, $\tilde{d}$ = 0.18.
{\it Left:} Purely CP-even, CP-odd or SM couplings.
{\it Right:} CP-even and CP-odd couplings of the same size.
}
\end{figure}

In order to determine the tensor structure of the $HVV$ couplings, for any scalar
particle $H$ found at the LHC,  the distributions of the two tagging jets
are an important tool. However, most distributions, like the jet
transverse momentum, jet energy or dijet invariant mass, may depend strongly
on the form factors and, therefore, are hard to predict without specifying the
underlying model of new physics. An exception is the azimuthal angle
between the two tagging jets in the final state. The shape of 
$d\sigma /d|\Delta\phi_{jj}|$  is quite insensitive 
to form factor effects~\cite{Figy:2004pt} and it provides for an
excellent distinction between the three tensor structures of
Eq.~(\ref{eq:vertex}).
The characteristic distributions are shown in
Fig.~\ref{fig:Azimuthal1}. For a purely CP-odd
coupling the cross section is suppressed at 0 and 180 degrees, for
a CP-even coupling this dip appears at 90
degrees, while a pure SM coupling produces a rather flat
$|\Delta\phi_{jj}|$ distribution~\cite{Plehn:2001nj}. Unfortunately,
when both CP-even and CP-odd couplings of similar strength are present,
the dips cancel and result in a distribution which is very similar to SM
expectations.

\begin{figure}[hbt]
\centerline{
\includegraphics[height=8cm, width=12cm]{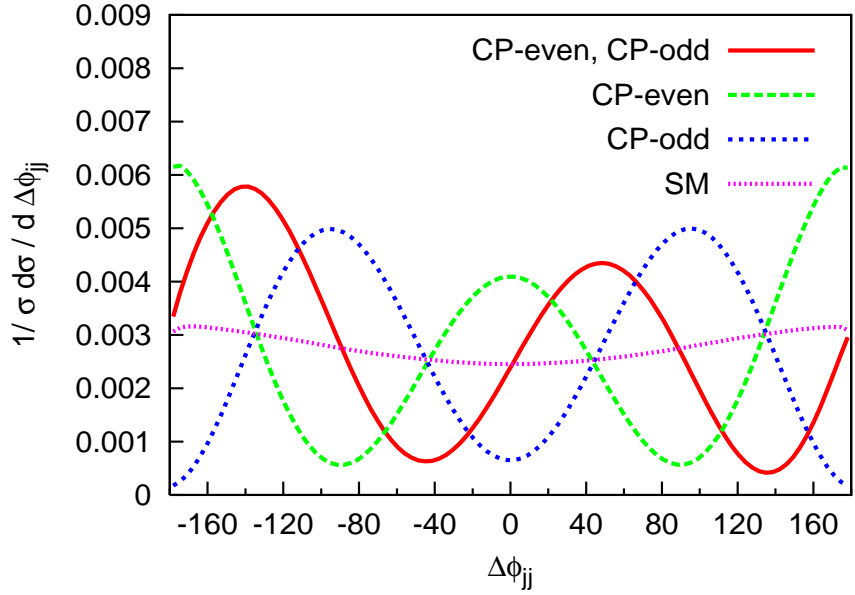}
}

\caption[]{\label{fig:Azimuthal2}
Normalized distribution of the azimuthal angle $\Delta\phi_{jj}$ defined
in Eq.~(\ref{eq:phijj}) for a Higgs mass of 120
GeV and a mixed CP scenario ($d$ = $\tilde{d}$ = 0.18, red solid curve),
a CP-even anomalous coupling ($d$ = 0.18, $\tilde{d}$ = 0, green dashed
curve), a CP-odd coupling ($d$ = 0, $\tilde{d}$ = 0.18, blue dotted
curve) and the SM case (purple narrow dotted line). }
\end{figure}

The missing information is contained in the sign of the azimuthal angle
between the tagging jets. Naively one might assume that this sign cannot
be defined unambiguously in $pp$ collisions because an azimuthal angle
switches sign when viewed along the
opposite beam direction. However, in doing so, the ``toward'' and the
``away'' tagging jets also switch place, i.e. one should take into
account the correlation of the tagging jets with the two distinct beam
directions.  Defining $\Delta\phi_{jj}$ as
the azimuthal angle of the ``away'' jet minus the azimuthal angle of the
``toward'' jet, a switch of the two beam directions leaves the sign of 
$\Delta\phi_{jj}$ intact. In order to be precise, let us
define the normalized four-momenta of the two proton beams as $b_+$ and
$b_-$, while $p_+$ and $p_-$ denote the 
four-momenta of the two tagging jets, where ${\bf p_+}$ points into the
same detector hemisphere as ${\bf b_+}$. Then  
\begin{equation} 
\label{eq:phijj}
\varepsilon_{\mu\nu\rho\sigma} b_+^\mu p_+^\nu b_-^\rho p_-^\sigma
= 2p_{T,+}p_{T,-}\sin(\phi_+ - \phi_-) = 2p_{T,+}p_{T,-}\sin\Delta\phi_{jj}
\end{equation} 
provides the sign of $\Delta\phi_{jj}$. This definition is manifestly
invariant under the interchange $(b_+,p_+)\leftrightarrow (b_-,p_-)$ and
we also note that $\Delta\phi_{jj}$ is a parity odd observable.

\begin{figure}[tb]
\centerline{
\includegraphics[height=7.2cm, width=7.8cm]{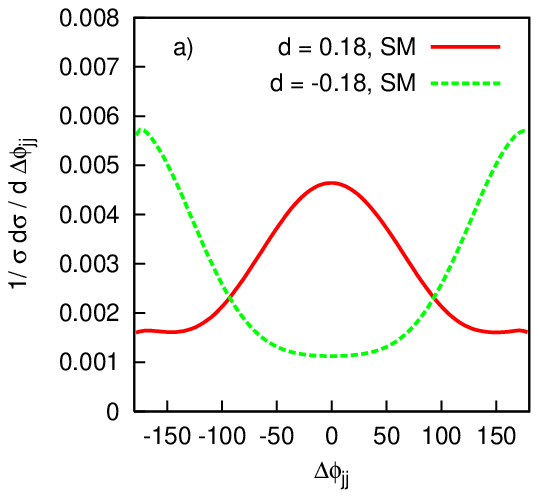} \hspace*{0.2cm}
\includegraphics[height=7.2cm, width=7.8cm]{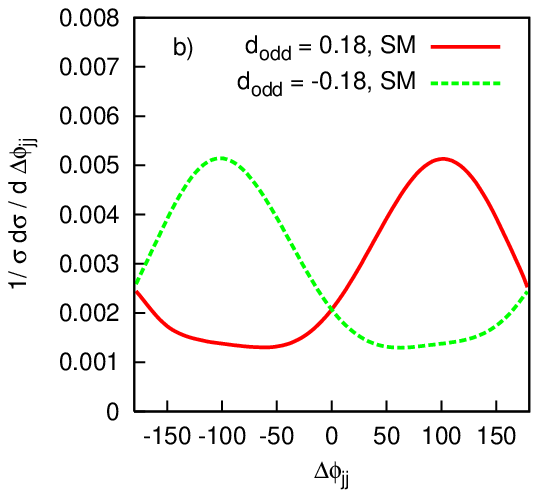}
}

\caption[]{\label{fig:deltaphialt2}
Normalized distribution of the azimuthal angle for a Higgs mass of 120
GeV for SM and anomalous couplings (a) $d=\pm 0.18$ and 
(b) $\tilde d = \pm 0.18$. The new definition of the
azimuthal angle is used.}
\end{figure}

The corresponding azimuthal angle distribution is shown in
Fig.~\ref{fig:Azimuthal2} for three scenarios of purely anomalous
couplings (i.e. no SM contribution, $a_1(q_1,q_2)=0$) and for the SM
case. The characteristic dips for purely CP-even and CP-odd couplings 
are still present and occur at $\pm$90 degrees and $0,\pm 180$ degrees,
respectively. In the case of mixed CP-even and CP-odd couplings,
\begin{equation}
\label{eq:mixedCP}
d = d_0\; \sin\alpha\;, \qquad \tilde d = d_0\; \cos\alpha\;, 
\end{equation} 
and no SM contribution, 
the positions of the dips shift to $\Delta\phi_{jj} = -\alpha$ and
$\Delta\phi_{jj} = -\alpha+\pi$ (modulo $2\pi$).
This shift also explains why $|\Delta\phi_{jj}|$ loses information in
the mixed CP case: when folding over the $\Delta\phi_{jj}$-distribution at
$\Delta\phi_{jj}=0$, the positions of the dips do not match and, hence,
they fill up.


A more complicated picture emerges when considering interference effects
between a SM contribution and anomalous couplings, i.e. when
$a_1$ and one or two of the anomalous form factors $a_i(q_1,q_2),\;
(i=2,3)$ of Eq.~(\ref{eq:vertex}) are present simultaneously. For
the sake of the argument, let us assume that $a_1$ has SM strength and
that $a_2$ and $a_3$ are real.
The full amplitude can then be written as 
\begin{equation}
M = M_{SM} + a_2 M_{CP-even} + a_3 M_{CP-odd}
\end{equation}
which results in the matrix element squared:
\begin{eqnarray}
\label{eq:Matrixel}
|M|^2 &=& |M_{SM}|^2 + a_2^2 |M_{CP-even}|^2 + a_3^2 |M_{CP-odd}|^2 
\nonumber \\
&&+
\underbrace{a_2 2Re\left( M_{SM}^*M_{CP-even}\right) + 
a_3 2Re\left( (M_{SM}+a_2M_{CP-even})^*M_{CP-odd}\right)
}_{\mathrm{Interference} \  \mathrm{term}}
\end{eqnarray}

For smallish anomalous couplings $a_2$ and/or $a_3$ the interference
terms provide the best sensitivity to new physics effects and these
interference terms also reveal the sign of anomalous couplings. 

When integrating over phase space, the three contributions $M_{SM}$,
$M_{CP-even}$ and $M_{CP-odd}$ may change relative signs and, thus,
interference effects may cancel. For the $|\Delta\phi_{jj}|$
distribution this is indeed a problem: $M_{SM}$ and $M_{CP-even}$ are
even functions of $\Delta\phi_{jj}$ while $M_{CP-odd}$ is odd in
$\Delta\phi_{jj}$. This means that the interference proportional to
$a_3$ integrates to zero when the sign of $\Delta\phi_{jj}$ is not
measured. The $a_2$ interference term, on the other hand, does not
suffer from this problem and can be observed in
$d\sigma/d|\Delta\phi_{jj}|$ already~\cite{Plehn:2001nj}: the shape of
the distribution changes dramatically when flipping the sign of the
anomalous coupling. 

The problem for the $a_3$ interference term is resolved by taking into
account the parity odd nature of the term linear in $M_{CP-odd}$, namely
by plotting $d\sigma/d\Delta\phi_{jj}$. The effect is demonstrated in 
Fig.~\ref{fig:deltaphialt2} where the interference of the SM amplitude
with a purely CP-even or purely CP-odd anomalous coupling is shown for
the two signs of the anomalous couplings. For the CP-even coupling the
interference term is even in $\Delta\phi_{jj}$ and, hence, is fully
present when plotting the absolute value of $\Delta\phi_{jj}$. For a
CP-odd anomalous coupling the interference is odd in $\Delta\phi_{jj}$
and completely disappears in $d\sigma/d|\Delta\phi_{jj}|$.

As noted above, $\Delta\phi_{jj}$ is a parity odd observable. Finding a 
$\Delta\phi_{jj}$ asymmetry as in Fig.~\ref{fig:deltaphialt2}(b) would show
that parity is violated in the process $qq\to qqH$. Since the SM
distribution (at tree level) is  $\Delta\phi_{jj}$-even, the parity
violation must originate from a parity-odd coupling, namely $a_3$ 
in the $HVV$ vertex. This term is also CP-odd. Such a coupling, occurring
at the same time as the CP-even SM 
amplitude or the CP-even coupling $a_2$, implies CP-violation in the
Higgs sector. In this sense, the observation of an asymmetry in the 
$\Delta\phi_{jj}$ distribution would directly demonstrate CP-violation
in the Higgs sector.

%
\section{Conclusions}
\label{sec:concl}

With VBFNLO we have available a parton level Monte Carlo program which
allows to calculate Higgs production via vector boson fusion, including
NLO QCD corrections. The program has now been extended to support
anomalous $HVV$ couplings in both the production and the Higgs decay
process. 

For Higgs masses below about 160 GeV, the non-observation of
$H\gamma$ signals at LEP puts stringent constraints on anomalous $HVV$
couplings. We have analyzed the size of deviations in VBF Higgs
signals due to anomalous couplings which are allowed by the LEP data.
Within these bounds, anomalous HWW couplings can enhance the
production cross section of a Higgs boson in VBF at the LHC by more than
10\% whereas effects from anomalous $H\gamma\gamma$ couplings are
negligible and the effects of anomalous $HZ\gamma$ and $HZZ$ couplings
are also small.
In contrast, the relevant anomalous
couplings in Higgs decay are H$\gamma\gamma$ and HZ$\gamma$. Even with the
existing bounds from LEP those couplings can enhance partial Higgs decay
widths by several orders of magnitude and, therefore, 
lead to measurable effects
in Higgs signals at the LHC.

A sensitive probe of anomalous Higgs couplings in the production process
is the azimuthal angle between the tagging jets, as defined in
Eq.~(\ref{eq:phijj}).  The azimuthal angle distribution is largely form
factor independent and allows to extract the complete 
information about the tensor structure of the $HVV$ coupling.
An odd contribution to the $\Delta\phi_{jj}$ distribution proves the
presence of parity violation in Higgs production and signals
CP-violation in the Higgs sector.

The $\Delta\phi_{jj}$ distribution should be analyzed for all $Hjj$
production processes at the LHC, in particular also for gluon
fusion. Since heavy fermion loops give rise to effective $Hgg$ vertices 
with the tensor structures given by the $a_2$ and $a_3$ terms in
Eq.~(\ref{eq:vertex}), the same qualitative behavior of
$d\sigma/d\Delta\phi_{jj}$ arises in gluon fusion events also and
distinguishes scalar and pseudo-scalar Higgs couplings to heavy
fermions~\cite{Hankele:2006ja} and also signals mixed scalar and
pseudo-scalar couplings, i.e. CP violating effects in the Higgs sector.
We will analyze these effects in a future publication~\cite{Klaemke}. 
 
%
%
\section*{Acknowledgments}
This research was supported by the Deutsche
Forschungsgemeinschaft in the Sonderforschungsbereich/Transregio 
SFB/TR-9 ``Computational Particle Physics'' and in the Gradu\-iertenkolleg
``High Energy Physics and Particle Astrophysics''.
%
%


\end{document}